\documentclass[preliminary,copyright,creativecommons]{eptcs}
\providecommand{\event}{ICLP 2026 Technical Communication} % Name of the event you are submitting to

\usepackage{hyperref}
\usepackage{iftex}

\ifpdf
  \usepackage{underscore}         % Only needed if you use pdflatex.
  \usepackage[T1]{fontenc}        % Recommended with pdflatex
\else
  \usepackage{breakurl}           % Not needed if you use pdflatex only.
\fi

\usepackage{amsmath}
\usepackage{graphicx}
\usepackage{multirow}
\usepackage{alltt}
\usepackage{amssymb}
\usepackage{tcolorbox}
\usepackage{esvect}
\usepackage{url}
\usepackage{plain}
\usepackage{float}
 \usepackage{tikz}
 \usetikzlibrary{arrows.meta}
 
\title{Case study: proving $\sqrt 2$ irrational with LPTP and an LLM}
\author{Fred Mesnard \qquad \'Etienne Payet
\institute{LIM, universit\'e de La R\'eunion, France}
\email{\{frederic.mesnard,etienne.payet\}@univ-reunion.fr}
\and Wim Vanhoof
\institute{Universit\'e de Namur, Belgium}
\email{wim.vanhoof@unamur.be}
}

\newcommand{\ie}{\textit{i.e.}, }
\newcommand{\eg}{\textit{e.g.}, }
\newcommand{\nat}{\mathbb{N}}
\newcommand{\DP}{DP} % Directly proven by Claude
\newcommand{\NP}{NP} % Not proven by Claude
\newcommand{\BF}{BF} % Proven by Claude with back and forths
\newcommand{\PG}{PG} % Proof given to Claude
\newcommand{\PH}{PH} % Proven by Claude with given hints

\newcommand{\titlerunning}{Case study: proving $\sqrt 2$ irrational with LPTP and an LLM}
\newcommand{\authorrunning}{F. Mesnard, \'E. Payet, W. Vanhoof}

\hypersetup{
  bookmarksnumbered,
  pdftitle    = {\titlerunning},
  pdfauthor   = {\authorrunning},
  pdfsubject  = {EPTCS},               
  pdfkeywords = {Prolog, Verification, LLM} % Uncomment and enter keywords specific to your paper
}

\begin{document}

\maketitle

\begin{abstract}
%Summary of the paper in about 250 words.
We present the interactions with an LLM (Large Language Model)
aiming at proving that $\sqrt 2$ is not a rational number
in an LP (Logic Programming) context.
We start from a few basic pure logic programming predicate definitions.
We rely on the LPTP  (Logic Program Theorem Prover) system written by 
Robert St{\"a}rk for stating and
proving properties about logic programs. 
As the proof language of LPTP  is based on natural deduction, 
the proofs are human readable.
In our case study, we sketch in LPTP the usual proof showing the irrationality of
$\sqrt 2$.  Then we describe the interactions we had with the LLM.
We end up with a complete formal
proof, partially generated by an LLM
and fully proof-checked by LPTP.

\end{abstract}

%\begin{keywords} Prolog  verification, LLM\end{keywords}

%\newpage
\section{Introduction}
\label{Introduction}
Consider the following pure Prolog program. 
Natural numbers are represented as terms
built from the constant $0$ and the function symbol $s/1$.
We borrow the predicates \texttt{nat/1}, \texttt{plus/3}, \texttt{times/3},
\texttt{gcd/3}, and \texttt{gcd\_leq/3} from the LPTP 
(Logic Program Theorem Prover,
\cite{Staerk98a}) libraries $\mathit{nat}$ and $\mathit{gcd}$
to facilitate the future use of  already proven properties.
\begin{verbatim}
even(0).                            odd(s(0)).
even(s(s(X))) :- even(X).           odd(s(s(X))) :- odd(X).

nat(0).                             leq(0,X).
nat(s(X)) :- nat(X).                leq(s(X),s(Y)) :- leq(X,Y).

plus(0,Y,Y).                        times(0,Y,0).
plus(s(X),Y,s(Z)) :- plus(X,Y,Z).   times(s(X),Y,Z) :- times(X,Y,P), plus(P,Y,Z).

gcd(X,Y,D) :-                       gcd_leq(X,Y,D) :-
 (	leq(X,Y) -> gcd_leq(X,Y,D)        (	X = 0 -> D = Y
  ;	gcd_leq(Y,X,D)).                  ;	plus(X,Z,Y), gcd(X,Z,D)).                       
\end{verbatim}  
\vspace{0.25cm}

\emph{Question:} can we find two coprime natural  numbers $p$ and $q$ such that
$ \frac{p}{q} = \sqrt 2$, \ie $p^2 = 2 q^2$? \\
(Two natural numbers $p$ and $q$ are \emph{coprime} if and only if their greatest common divisor is 1.)
This question
corresponds  to the following Prolog query, designed so that 
backtracking generates a fair 
exploration of $\nat \times \nat$, where
$\nat$ denotes the set of natural numbers:

\begin{verbatim}
?- nat(S),plus(P,Q,S),gcd(P,Q,s(0)),times(P,P,P2),times(Q,Q,Q2),plus(Q2,Q2,P2).
\end{verbatim}

The answer to the question  is \emph{no}.
This fact is known since at least the 6th century BC.
But since the search space of the above query is infinite, we won't get a negative
answer from \emph{any} resolution-based logic programming (LP) engine, Prolog included.
However we can prove this fact by other means, still within an LP
context. For instance,
we can prove $\sqrt 2$ is not rational using
the LPTP system.
We take profit of this exercise to evaluate the help we may get from an LLM (Large Language Model).

Recent research has shown that LLMs can be made increasingly effective at automated theorem proving by integrating them with formal provers and proof checkers (see, \eg \cite{LPAR2024:Automated-Theorem-Provers-Help,rao2025neuraltheoremprovinggenerating}). To the best of our knowledge, no such integration has been attempted for LPTP. Yet, LPTP is an interesting formalism and tool as it uses Prolog as representation language, and the associated prover constitutes a rather lightweight and easily manipulable process. 
The theoretical basis and specification
language is purely first-order logic, 
which is well-known and largely automatable.
% \textbf{Est-ce correct, voyez-vous d'autres avantages?} 
On the other hand, less documentation and examples might be available in comparison with other provers (such as Rocq, Lean and Isabelle), which might make it more difficult for an LLM to appropriate the formalism. %Hence, our experiment.

The paper is organized as follows. Section~\ref{A-quick-summary-of-LPTP}
recalls the basics of LPTP.
Section~\ref{Our-initial-draft} presents 
a usual proof of the irrationality of $\sqrt 2$
and a formalization of its skeleton.
Section~\ref{The-proof-completed} and~\ref{Back-to-Prolog} 
explain how we interact with the LLM to get a complete
proof. Section~\ref{Related-work} reviews some related work, 
and Section~\ref{Conclusion} concludes.

\section{A quick summary of LPTP}
\label{A-quick-summary-of-LPTP}

The reader already familiar with LPTP can safely skip this section.
Let $P$ be a pure logic program where negative literals
may appear in the body of clauses (also called \emph{normal  program} in  \cite{Lloyd87a}).
For sake of conciseness,
we do not consider built-in predicates (see \cite{Staerk98a}
for a full treatment) other than the equality \texttt{=/2}.
We start with $\cal{L}$, the first-order language associated to $P$. 
The \emph{goals} of $\cal{L}$ are:
$$G,H ::= \texttt{true} \ | \texttt{fail}\ | \ s = t \ | \ A \ | \ \texttt{\textbackslash +} \ G \ | 
     \ (G,H) \  | \ (G;H) \ | \ \texttt{some} \ x \ G$$
where $s$ and $t$ are two terms, $x$ is a variable and $A$ is an atomic goal. The goals of $\cal{L}$ have the operational semantics specified 
by ISO-Prolog \cite{ISO-Prolog95} assuming the occurs check.

$\hat{\cal{L}}$ is the specification language of LPTP.
For each user-defined predicate symbol $R$,  $\hat{\cal{L}}$ does not include $R$, but instead it contains
three predicate symbols $R^s$, $R^f$, $R^t$ of the same arity as $R$,
which respectively express success, failure and termination of $R$. 
$\hat{\cal{L}}$ also contains a unary  constraint for groundness $gr$, expressing that its argument is ground.
The \emph{formulas} of $\hat{\cal{L}}$ are:
$$\phi,  \psi ::= \top \ | \ \bot \ | \ s = t  \ | \ R(\vv{t}) \ | 
  \  \neg \phi \ |  \ \phi \land \psi \ |  \ \phi \lor \psi \ |  \ \phi \rightarrow \psi \ | \ \forall x \phi \ | \ \exists x \phi $$
where $\vv{t}$ is a sequence of  $n$ terms and $R$ denotes a $n$-ary predicate symbol of $\hat{\cal{L}}$. The semantics of $\hat{\cal{L}}$ 
is the first-order predicate calculus of classical logic.

For any of the user-defined  logic procedures $R$ in a logic program $P$,
$D^P_R(\vv{x})$ denotes its Clark's \emph{if-and-only-if} completed definition, c.f. \cite{Cla78, Lloyd87a}.

For defining the declarative semantics of logic programs, 
St{\"a}rk uses three syntactic operators \textbf{S}, \textbf{F} and \textbf{T} which map
goals of $\cal{L}$ into $\hat{\cal{L}}$-formulas. Intuitively, $\textbf{S}G$ means $G$ succeeds
(any breadth-first evaluation of $G$ succeeds), 
$\textbf{F}G$ means $G$  fails (the ISO-Prolog evaluation stops without any answer), 
and $ \textbf{T}G$ means $G$ terminates
(the ISO-Prolog evaluation produces a finite number of answers then stops).
Moreover, termination implies a safe use of negation.
The definition of the operators follows:
\begin{flushleft}
\small
\begin{tabular}{l l l l}
%\hline
%%
 $\textbf{S}R(\vv{t}) := R^s(\vv{t})$ 	&   
 	$\textbf{S}  \ \texttt{true} := \top$ & $\textbf{S} \ \texttt{fail} := \bot$ & $\textbf{S} (s = t) := (s = t)$ \\
 $\textbf{S}  \texttt{\textbackslash +} G := \textbf{F} G$ & $\textbf{S}( G, H) := \textbf{S}G \land \textbf{S}H$  &
 	$\textbf{S}( G; H) := \textbf{S}G \lor \textbf{S}H$  & $\textbf{S}(\texttt{some} \ x \ G) := \exists x \textbf{S}G$  \\
	& & & \\
 $\textbf{F}R(\vv{t}) := R^f(\vv{t})$ 	&   
 	$\textbf{F}  \ \texttt{true} := \bot$ & $\textbf{F} \ \texttt{fail} := \top$ & $\textbf{F} (s = t) := \neg (s = t)$ \\
 $\textbf{F}  \texttt{\textbackslash +} G := \textbf{S} G$ & $\textbf{F}( G, H) := \textbf{F}G \lor \textbf{F}H$  &
 	$\textbf{F}( G; H) := \textbf{F}G \land \textbf{F}H$  & $\textbf{F}(\texttt{some} \ x \ G) := \forall x \textbf{F}G$  \\
%%
%	& & & \\
% $\textbf{T}R(\vv{t}) := R^t(\vv{t})$ 	&   
 %	$\textbf{T}  \ \texttt{true} := \top$ & $\textbf{T} \ \texttt{fail} := \top$ & $\textbf{T} (s = t) := \top$ \\
%%
 %$\textbf{T}  \texttt{\textbackslash +} G := \textbf{T} G\land  gr(G)$ & &
 %	$\textbf{T}( G, H) := \textbf{T}G \land( \textbf{F}G \lor \textbf{T}H)$  & \\
% 	$\textbf{F}( G; H) := \textbf{T}G \land \textbf{T}H$  & $\textbf{T}(\texttt{some} X \ G) := \forall x \textbf{T}G$  \\
%%	
%\hline
\end{tabular}
\end{flushleft}

\vspace{0.1cm}
\begin{flushleft}
\small
\begin{tabular}{l l }
%\hline
 $\textbf{T}R(\vv{t}) := R^t(\vv{t})$ 	&   
 	$\textbf{T}  \ \texttt{true} := \top$  \\
$\textbf{T} \ \texttt{fail} := \top$ &  $\textbf{T} (s = t) := \top$  \\
 $\textbf{T}  \texttt{\textbackslash +} G := \textbf{T} G\land  gr(G)$ & 
 	$\textbf{T}( G, H) := \textbf{T}G \land( \textbf{F}G \lor \textbf{T}H)$   \\
$\textbf{T}( G; H) := \textbf{T}G \land \textbf{T}H$  &  $\textbf{T}(\texttt{some} \ x \ G) := \forall x \textbf{T}G$   \\
%%	
%\hline
\end{tabular}
\end{flushleft}

With LPTP, we prove properties of a logic program $P$ w.r.t. its \emph{inductive extension} IND($P$)
which includes Clark's completion \cite{Cla78} and induction along the definition
of the predicates. St{\"a}rk shows that the inductive extension is always consistent
and proves various correctness and completeness results w.r.t. the operational semantics of Prolog \cite{Staerk98a}.
The first-order theory IND($P$) (cf. \cite{Staerk98a}, pp. 253--254) is defined by 
nine axiom schemas
which we describe now.
We omit the fixed point axioms for builtins.  
Let us point out that the specification language $\hat{\cal{L}}$ of LPTP can
be extended by new function and predicate symbols, 
which can be quite handy while formalizing properties.

The first two axioms specify some properties of the
trees built from the  function symbols extracted from the program under consideration. 
The third axiom forbids infinite %rational 
trees.
\begin{tcolorbox}[coltitle=black!75!black, colbacktitle=black!10!white,
                title=The axioms of Clark's equality theory]
\begin{itemize}
    \item[1.] $f(x_1,\ldots, x_n) = f(y_1,\ldots, y_n) \rightarrow x_i = y_i$ [if $f$ is $n$-ary and $1 \leq i \leq n$]
    \item[2.] $f(x_1,\ldots, x_n) \neq g(y_1,\ldots, y_m)$ [if $n \neq m$ or $f  \not\equiv g$]
    \item[3.] $t \neq x$ [if $x$ occurs in $t$ and $t  \not\equiv x$]
\end{itemize}
\end{tcolorbox}

LPTP deals with \emph{non-ground} terms, as any ISO-Prolog processor does.
LPTP offers a predefined predicate  $gr/1$ that we can consider as a constraint.
This relation is useful for instance for 
dealing with negation as failure as
LPTP only allows negation by failure for \emph{ground} goals
(see the  definition $\textbf{T}  \texttt{\textbackslash +} G$).
\begin{tcolorbox}[coltitle=black!75!black, colbacktitle=black!10!white,
                title=Axioms for gr/1]
\begin{itemize}
    \item[4.] gr($c$) [if $c$ is a constant]
    \item[5.] $\textup{gr}(x_1) \land \ldots \land  \textup{gr}(x_m) \leftrightarrow \textup{gr}(f(x_1,\ldots,x_m))$ [$f$ is $m$-ary]
\end{itemize}
\end{tcolorbox}

LPTP considers each user-defined predicate through three points of view: failure, success and termination.
These three viewpoints are linked with the following axioms.
%where $R^s$ (resp. $R^f$ and $R^t$) denotes 
%\texttt{R\_succeeds/3} (resp. \texttt{R\_fails/3} and \texttt{R\_terminates/3}). 

%
\begin{tcolorbox}[coltitle=black!75!black, colbacktitle=black!10!white,
                title=Uniqueness axioms and totality axioms]
\begin{itemize}
    \item[6.] $\neg (R^s(\vv{x}) \land R^f(\vv{x}))$ [if $R$ is a user-defined predicate]
    \item[7.] $R^t(\vv{x}) \rightarrow (R^s(\vv{x}) \lor R^f(\vv{x}))$ [if $R$ is a user-defined predicate]
\end{itemize}
\end{tcolorbox}
Axiom 6 says that for any tuple of (possibly non-ground)  terms, we cannot have at the same time success and failure of $R$.
Axiom 7 states that given termination, we have success or failure. Altogether, it means that for any 
tuple of terms $\vv{x}$, assuming termination,
either $R(\vv{x})$ succeeds or (exclusively) $R(\vv{x})$  fails.
\begin{tcolorbox}[coltitle=black!75!black, colbacktitle=black!10!white,
                title=Fixed point axioms for user-defined predicates $R$]
\begin{itemize}
    \item[8.] $R^s(\vv{x})  \leftrightarrow$ \textbf{S}$D^P_R(\vv{x})$, 
    \ $R^f(\vv{x})  \leftrightarrow$ \textbf{F}$D^P_R(\vv{x})$,
    \ $R^t(\vv{x})  \leftrightarrow$ \textbf{T}$D^P_R(\vv{x})$
\end{itemize}
\end{tcolorbox}
We recall that $D^P_R(\vv{x})$ denotes the definition of the completion \cite{Cla78} 
of the user-defined  procedure $R(\vv{x})$ in the logic program $P$.
In the previous section, we saw how to apply the operator \textbf{S}, \textbf{F} and \textbf{T}
to formulas. So for instance, the first equivalence $R^s(\vv{x})  \leftrightarrow$ \textbf{S}$D^P_R(\vv{x})$
defines $R^s(\vv{x})$. 

Finally, for any  property
of the form $\forall \vv{x} [R^s(\vv{x}) \rightarrow \phi(\vv{x})]$, where $R(\vv{x})$ 
is a user-defined procedure  and  $\phi(\vv{x})$ an $\hat{\cal{L}}$ -formula,
we have an induction schema.
%Given a logic program $P$ as a finite text, there is a finite number
%of user-defined procedures but an infinite number of formula. 
The interactive prover LPTP is able to \emph{dynamically} generate 
an induction axiom on demand while the user interacts with it.
Let us examine a simple case. It is exactly what happens using LPTP,
which slightly generalizes \cite{Staerk98a}.
By \emph{directly recursive  user-defined predicate} in the box below, 
we forbid mutual recursive definitions.
Of course, LPTP is able to handle mutually recursive properties, see \cite{Staerk95a} for some examples.
\begin{tcolorbox}[coltitle=black!75!black, colbacktitle=black!10!white,
                title=A (simplified) induction schema for a user-defined predicate $R$ ]
    Let $R$ be a directly recursive  user-defined predicate and 
    let $\phi(\vv{x})$ be an $\hat{\mathcal{L}}$-formula such that the length of $\vv{x}$ is equal to the arity of $R$.\\
    Let $sub(\phi(\vv{x})/R)$ be the formula to be proven
    $\forall \vv{x} (R^s(\vv{x}) \rightarrow \phi(\vv{x}))$. \\
    Let $closed(\phi(\vv{x})/R)$ be the formula obtained from 
    $\forall \vv{x} (\textbf{S} D^P_R(\vv{x}) \rightarrow  R^s(\vv{x}))$ by replacing 
    \begin{itemize}
    \item $R^s(\vv{x})$ by $\phi(\vv{x})$  	on the right of $\rightarrow$,
    \item all occurrences of $R^s(\vv{t})$ appearing on the left of $\rightarrow$ by $\phi(\vv{t}) \land R^s(\vv{t})$ (these
    are the recursive calls of $R$ in $D^P_R$, turned into $R^s$ by the operator \textbf{S}).
    \end{itemize}
    Then the induction axiom is the following formula:
\begin{itemize}
    \item[9.] $closed(\phi(\vv{x})/R) \rightarrow sub(\phi(\vv{x})/R)$
\end{itemize}
\end{tcolorbox}
%
%\textbf{Mettre un exemple?}\\
%
%
  As an illustration, consider the lemma
  $\forall x \ ( \mathit{even}^s(x) \rightarrow \mathit{odd}^s(s(x)))$, which we will use in
  section~\ref{The-proof-completed} (lemma \textit{even:succ:odd}).
  We have $\vv{x} = x$, $R \equiv \mathit{even}$ and $\phi(x) \equiv \mathit{odd}^s(s(x))$
  and $sub(\phi(x)/R)$ is precisely lemma \textit{even:succ:odd}.
  Now for $closed(\phi(x)/R)$,
  recall the clauses defining \textit{even} and \textit{odd}:
  \texttt{even(0)}, \texttt{even(s(s(X))) :- even(X)}, \texttt{odd(s(0))} and
  \texttt{odd(s(s(X))) :- odd(X)}.
  The Clark completion of \textit{even} is
  $D^P_{\mathit{even}}(x) \equiv x = 0 \lor \exists y\,(x = s(s(y)) \land \mathit{even}(y))$,
  and applying the operator \textbf{S} gives
  $\textbf{S}\,D^P_{\mathit{even}}(x) \equiv x = 0 \lor \exists y\,(x = s(s(y)) \land \mathit{even}^s(y))$.
  After the two
  substitutions ($\mathit{even}^s(x) \rightsquigarrow \phi(x)$ on the right and
  $\mathit{even}^s(y) \rightsquigarrow \phi(y) \land \mathit{even}^s(y)$ on the left), we get:
  $
  closed(\phi(x)/\mathit{even}) \;=\; \forall x\,\Bigl[\bigl(x = 0 \,\lor\, \exists y\,(x = s(s(y)) \land \phi(y) \land \mathit{even}^s(y))\bigr) \rightarrow \phi(x)\Bigr]
  $.
Let us simplify this formula using three standard equivalences of first-order logic: (a) $\forall x\,[(A \lor B) \rightarrow C] \;\equiv\; \forall x\,(A \rightarrow C) \,\land\, \forall x\,(B \rightarrow C)$; (b) $\forall x\,(x = t \rightarrow \psi(x)) \;\equiv\; \psi(t)$ (when $x$ is not free in~$t$);
(c) $\forall x\,[\exists y\,P(x,y) \rightarrow Q(x)] \;\equiv\; \forall x\,\forall y\,[P(x,y) \rightarrow Q(x)]$ (when $y$ is not free in~$Q$).
%  \begin{itemize}
%  \item[(a)] $\forall x\,[(A \lor B) \rightarrow C] \;\equiv\; \forall x\,(A \rightarrow C) \,\land\, \forall x\,(B \rightarrow C)$;
%  \item[(b)] the one-point rule $\forall x\,(x = t \rightarrow \psi(x)) \;\equiv\; \psi(t)$ (when $x$ is not free in~$t$);
%  \item[(c)] $\forall x\,[\exists y\,P(x,y) \rightarrow Q(x)] \;\equiv\; \forall x\,\forall y\,[P(x,y) \rightarrow Q(x)]$ (when $y$ is not free in~$Q$).
%  \end{itemize}
  Applying~(a) splits $closed(\phi(x)/\mathit{even})$ into two conjuncts.
  The first one, $\forall x\,(x = 0 \rightarrow \phi(x))$, reduces to $\phi(0)$ by~(b).
  The second one, $\forall x\,[\exists y\,(x = s(s(y)) \land \phi(y) \land \mathit{even}^s(y)) \rightarrow \phi(x)]$,
  becomes after extraction of the existential by~(c) and elimination of~$x$ by~(b) (using $x = s(s(y))$):
  $\forall y\,[\phi(y) \land \mathit{even}^s(y) \rightarrow \phi(s(s(y)))]$.
  Renaming $y$ to $x$ and combining with $sub(\phi(x)/\mathit{even}) = \forall x\,(\mathit{even}^s(x) \rightarrow \phi(x))$,
  the induction axiom $closed(\phi(x)/\mathit{even}) \rightarrow sub(\phi(x)/\mathit{even})$ instantiates to:
% $$
%  \bigl[\phi(0) \,\land\, \forall x\bigl(\phi(x) \land \mathit{even}^s(x) \rightarrow
%  \phi(s(s(x)))\bigr)\bigr]
%  \;\rightarrow\; \forall x\bigl(\mathit{even}^s(x) %\rightarrow \phi(x)\bigr)
%  $$
%  which gives the induction axiom for the initial lemma:
  $
  \bigl[\mathit{odd}^s(s(0)) \,\land\, \forall x\bigl(\mathit{odd}^s(s(x)) \land \mathit{even}^s(x) \rightarrow
  \mathit{odd}^s(s(s(s(x))))\bigr)\bigr]
  \;\rightarrow\; \forall x\bigl(\mathit{even}^s(x) \rightarrow \mathit{odd}^s(s(x))\bigr).
  $
% Let us have a look at the premise.
% The base case $\mathit{odd}^s(s(0))$ holds by the clause
% \texttt{odd(s(0))}.
% The inductive step follows from the clause \texttt{odd(s(s(X))) :- odd(X)}
%  instantiated with $X = s(x)$: from $\mathit{odd}^s(s(x))$ we derive
%  $\mathit{odd}^s(s(s(s(x))))$. The induction  axiom allows us to
%  close the proof of lemma \textit{even:succ:odd}.
%  Note that the recursion step is by~$2$ rather than by~$1$:
%  the schema mirrors the structural recursion of the predicate on which
%  induction is performed.
%

We refer the reader to the papers of St{\"a}rk \cite{Staerk98a,Staerk96a,Staerk95a}
for a complete presentation of LPTP.

%\subsection{A TeX proof + only the statement}
%
%\begin{plain}
%\input{add_x_0_x.tex}	
%\end{plain}

\section{Our initial draft}
\label{Our-initial-draft}

Let us go back to our introductory example.
The informal reasoning proving the irrationality of $\sqrt 2$ 
that we start with is the following usual proof by contradiction.
\begin{quote}
Assume we have two coprime natural numbers $p$ and $q$
such that 
$$p^2 = 2 q^2$$  
So $p^2$ is even. Hence $p$ 
is even, because the square of an odd number is odd.
Thus there exists a natural number $r$ such that
$p = 2 r$. By replacing $p$ with $2 r$ in the main equation, we get
$4 r^2 = 2 q ^2$, which gives $2 r^2 = q^2$.
Hence $q$ is also even, which contradicts our assumption.
\end{quote}

We formalize the above reasoning by first introducing 
the logical relations $\mathit{divisor}/2$ (which we get from the LPTP library) and $\mathit{coprime}/2$. 
We also define the logical function $\mathit{square}/1$ which returns the square of its argument.
%The possibility of defining logical relations and functions
%is a very handy extension introduced in LPTP by St{\"a}rk.
%Logical relations are simply new predicates 
%acting as logical abbreviations.
%Logical functions extend the set of terms, allowing more concise
%expressions compared to a pure relational way of writing formulas. 
Before defining a function, we must show that the associated definition 
(here $n \mapsto p \text{ such that }  \mathit{times}(n,n,p)$ holds) is actually functional: for 
any $\mathit{nat}(n)$, there exists a unique $p$ verifying $\mathit{times}(n,n,p)$.
So we need to prove existence and uniqueness, as done below.
For proving these two properties, we rely on 
lemma \textit{times:existence}
and   lemma \textit{times:uniqueness},
already proven in the LPTP library \textit{nat}.

%\newpage
\begin{plain}
%   Author: Robert Staerk <staerk@math.stanford.edu>
%  Created: January 1995
%  Updated: Fri Mar 12 09:05:39 2004 
% Filename: proofmacros.tex
%
% Font for operators `succeeds', `fails' and `terminates'.
%
\font\bfsf=cmssbx10
%
% The style of proofs is "ragged right".
%
\rightskip 0pt plus 10cm
\tolerance=400
%\hoffset=-4.5mm
%
% Macros used in formulas:
%
%  \D          definition
%  \F          failure operator
%  \It          italic font for defined functions
%  \S          success operator
%  \T          termination operator
%  \Tt         typewriter font for predicates
%  \all        universal quantifier
%  \app        X ** Y
%  \eq         equality
%  \eqv        equivalence
%  \ex         existential quantifier
%  \is         X is Y
%  \land       conjunction
%  \leq        less than or equal to
%  \lnot       negation
%  \lor        disjunction
%  \neq        disequality
%  \sub        subset
%  \to         implication
%  \v          integer variables
%
\def\D{\mathop{\hbox{\bfsf D}}}
\def\F{\mathop{\hbox{\bfsf F}}}
\def\It#1{\hbox{\it #1}}
\def\S{\mathop{\hbox{\bfsf S}}}
\def\Tt#1{\hbox{\tt #1}}
\def\T{\mathop{\hbox{\bfsf T}}}
\def\all[#1]{\forall #1}
\def\app{\nobreak\mathbin{**}\nobreak}
\def\eqv{\leftrightarrow\penalty\levcount}
\def\eq{\nobreak=\nobreak}
\def\ex[#1]{\exists #1}
\def\is{\nobreak\mathbin{\hbox{\tt is}}\nobreak}
\def\land{\wedge\penalty\levcount}
\def\lor{\vee\penalty\levcount}
\def\sub{\nobreak\subseteq\nobreak}
\def\to{\rightarrow\penalty\levcount}
\def\apply{\nobreak\mathbin{/. }\nobreak}
\def\v#1{v_{#1}}
%
% The depth of formulas:
%
\newcount\levcount
\def\0{\global\levcount=20}
\def\1{\global\advance\levcount by 20}
\def\2{\global\advance\levcount by -20}
%
% Underscores (cf. ^ and ^^ in manmac.tex):
%
\newif\ifref
\reffalse
%\def\specialunderscore{\ifmmode_\else\ifref-\else\_\fi\fi}
%\catcode`\_=13 % active
%\let _=\specialunderscore
%
% Labels and backward references:
%
\def\label#1#2{\reftrue\expandafter\edef\csname#1:#2\endcsname{\Hlink{\number\thmcount}}%
\Htarget{\number\thmcount}%
\edef\next{\write\auxout{\string\newlabel{#1}{#2}{\jobname}{\number\thmcount}}}%
\next\reffalse\ignorespaces}
\def\by#1#2{\penalty 20\ by\penalty 20\ #1~\reftrue%
\expandafter\ifx\csname#1:#2\endcsname\relax#2%
\else\csname#1:#2\endcsname\fi\reffalse~[{\it #2}]}
\def\newlabel#1#2#3#4{\expandafter\edef\csname#1:#2\endcsname{#4 in Module {\tt #3}}}
%
% Hyperlinks (based on `hyperref.dtx' by Sebastian Rahtz):
%
\def\Hlink#1{#1}
\def\Htarget#1{}
%
% Uncomment the following lines if you want to use hyperlinks.
%
\def\Hend{}
\edef\Hhash{\string#}
\edef\Hquote{\string"}
\def\Hhref#1{}
\def\Hname#1{}
\def\Hlink#1{\Hhref{#1}#1\Hend}
\def\Htarget#1{\Hname{#1}\Hend}
%
% Theorem, Lemma, Corollary, Definition, Axiom:
%
\newcount\thmcount
\thmcount=0
\newcount\indcount
\def\inc{\global\advance\indcount by 1\hangindent\indcount em}
\def\dec{\global\advance\indcount by-1}
\def\nl{\par\hangindent\indcount em\noindent\kern\indcount em\ignorespaces} 
\def\lev{$\strut_{\number\indcount}$}
\def\Module#1#2#3{\bigskip\goodbreak % \vfil\allowbreak\vfilneg
  \advance\thmcount by1\label{#1}{#2}
%  \noindent{\bf #1~\the\thmcount}~[{\it #2}] $#3$.\allowbreak}
  \noindent{\bf #1}~[{\it #2}] $#3$.\allowbreak}
\def\theorem{\Module{Theorem}}
\def\lemma{\Module{Lemma}}
\def\corollary{\Module{Corollary}}
\def\definition{\Module{Definition}}
\def\axiom{\Module{Axiom}}
%
% Proof steps:
%
\def\Pr{\allowbreak\smallskip\noindent\global\indcount=0{\bf Proof. }\nobreak}
\def\Epr{\hbox{\rlap{$\sqcup$}$\sqcap$}\smallskip\allowbreak}
\def\Ass#1{\nl Assumption\lev: $#1$. \inc}
\def\Eass#1{\dec\nl Thus\lev: $#1$.}
\def\Cas#1{\nl Case\lev: $#1$. \inc}
\def\Ecas{\dec}
\def\Fin#1{\nl Hence\lev, in all cases: $#1$.}
\def\Dir#1{\nl Indirect\lev: $#1$. \inc}
\def\Edir#1{\dec\nl Thus\lev: $#1$.}
\def\Con#1{\nl Contra\lev: $#1$. \inc}
\def\Econ#1{\dec\nl Thus\lev: $#1$.}
\def\Ex[#1]#2{\nl Let\lev\ $#1$ with $#2$. \inc}
\def\Eex#1{\dec\nl Thus\lev: $#1$.}
\def\Ind#1{\nl Induction\lev: #1. \inc}
\def\Eind{\dec}
\def\Stp#1{\nl Hypothesis\lev: #1. \inc}
\def\Estp#1{\dec\nl Conclusion\lev: $#1$.}
\def\noproofs{\let\Pr=\nil\let\Epr=\par}
\def\nil#1{}
% FM
%\def\title#1{\noindent{\bf File:} {\tt#1.pr}\par}
%
% .aux
%
\newread\inputcheck
\def\inputaux#1{\openin\inputcheck #1 \ifeof\inputcheck \message
  {No file #1.}\else\closein\inputcheck \relax\input #1 \fi}
\newwrite\auxout
\openout\auxout=\jobname.aux
\let\endsave=\end
\def\end{\write\auxout{}\closeout\auxout\endsave}
%
%
% "!" is escape character like backslash "\"
%
\catcode `!=0
\catcode `@=11
\catcode `#=11
\catcode `&=11
%
%\noproofs
%
%
% End of proofmacros.tex

%!title{coprime_square}

!definition{divisor/2}{!0!1!all[x,y]!,!1(!It{divisor}(x,y)!eqv !1!ex[z]!,!1(!1!S !Tt{nat}(z)!2!land x*z !eq y)!2!2)!2!2}

!definition{coprime/2}{!0!1!all[p,q]!,!1(!It{coprime}(p,q)!eqv !1!all[r]!,!1(!1!1!S !Tt{nat}(r)!2!land !It{divisor}(r,p)!land !It{divisor}(r,q)!2!to r !eq !Tt{s}(!Tt{0}))!2!2)!2!2}

!lemma{square:existence}{!0!1!all[n]!,!1(!1!S !Tt{nat}(n)!2!to !1!ex[p]!,!1!S !Tt{times}(n,n,p)!2!2)!2!2}
!Pr{
!Ass{!0!1!S !Tt{nat}(n)!2}
$!0!1!1!S !Tt{nat}(n)!2!land !1!S !Tt{nat}(n)!2!2$.
$!0!1!ex[z]!,!1!S !Tt{times}(n,n,z)!2!2$!by{Lemma}{times:existence}.
!Eass{!0!1!1!S !Tt{nat}(n)!2!to !1!ex[p]!,!1!S !Tt{times}(n,n,p)!2!2!2}}
!Epr

!lemma{square:uniqueness}{!0!1!all[n,sn,sp]!,!1(!1!1!S !Tt{times}(n,n,sn)!2!land !1!S !Tt{times}(n,n,sp)!2!2!to sn !eq sp)!2!2}
!Pr{
!Ass{!0!1!1!S !Tt{times}(n,n,sn)!2!land !1!S !Tt{times}(n,n,sp)!2!2}
$!0sn !eq sp$!by{Lemma}{times:uniqueness}.
!Eass{!0!1!1!1!S !Tt{times}(n,n,sn)!2!land !1!S !Tt{times}(n,n,sp)!2!2!to sn !eq sp!2}}
!Epr

!definition{square/1}{!0!1!all[n,p]!,!1(!1!S !Tt{nat}(n)!2!to !1(!It{square}(n) !eq p!eqv !1!S !Tt{times}(n,n,p)!2)!2)!2!2}

!end
	
\end{plain}
\vspace{0.25cm}

%\textbf{Expliquer lemmes biblio}\\
Now we proceed top-down, starting from the informal proof
which we translate in LPTP as our main theorem.
We also need auxiliary results, and
we ask to the proof checker to admit them by using the LPTP tactic
\emph{by gap}. The proof checker validates the whole proof skeleton
and produces a \TeX file. Here is its PDF rendering.

\begin{plain}

!theorem{sqrt2:irrational}{!0!1!all[p,q]!,!1(!1!1!S !Tt{nat}(p)!2!land !1!S !Tt{nat}(q)!2!land !It{coprime}(p,q)!2!to !1!lnot !It{square}(p) !eq !Tt{s}(!Tt{s}(!Tt{0}))*!It{square}(q)!2)!2!2}
!Pr{
!Ass{!0!1!1!S !Tt{nat}(p)!2!land !1!S !Tt{nat}(q)!2!land !It{coprime}(p,q)!2}
!Con{!0!It{square}(p) !eq !Tt{s}(!Tt{s}(!Tt{0}))*!It{square}(q)}
$!0!1!S !Tt{even}(!It{square}(p))!2$! by~{!bf GAP}.
$!0!1!S !Tt{even}(p)!2$! by~{!bf GAP}.
$!0!1!ex[r]!,!1(!1!S !Tt{nat}(r)!2!land p !eq !Tt{s}(!Tt{s}(!Tt{0}))*r)!2!2$! by~{!bf GAP}.
!Ex[r]{!0!1!1!S !Tt{nat}(r)!2!land p !eq !Tt{s}(!Tt{s}(!Tt{0}))*r!2}
$!0!It{square}(p) !eq !It{square}(!Tt{s}(!Tt{s}(!Tt{0})))*!It{square}(r)$! by~{!bf GAP}.
$!0!Tt{s}(!Tt{s}(!Tt{s}(!Tt{s}(!Tt{0}))))*!It{square}(r) !eq !Tt{s}(!Tt{s}(!Tt{0}))*!It{square}(q)$! by~{!bf GAP}.
$!0!Tt{s}(!Tt{s}(!Tt{0}))*!It{square}(r) !eq !It{square}(q)$! by~{!bf GAP}.
$!0!1!S !Tt{even}(q)!2$! by~{!bf GAP}.
!Eex{!0!1!S !Tt{even}(q)!2}
$!0!It{divisor}(!Tt{s}(!Tt{s}(!Tt{0})),q)$! by~{!bf GAP}.
$!0!It{divisor}(!Tt{s}(!Tt{s}(!Tt{0})),p)$! by~{!bf GAP}.
$!0!1!S !Tt{nat}(!Tt{s}(!Tt{s}(!Tt{0})))!2$.
$!0!It{coprime}(p,q)$.
$!0!1!all[r]!,!1(!1!1!S !Tt{nat}(r)!2!land !It{divisor}(r,p)!land !It{divisor}(r,q)!2!to r !eq !Tt{s}(!Tt{0}))!2!2$!by{Definition}{coprime/2}.
$!0!1!1!1!S !Tt{nat}(!Tt{s}(!Tt{s}(!Tt{0})))!2!land !It{divisor}(!Tt{s}(!Tt{s}(!Tt{0})),p)!land !It{divisor}(!Tt{s}(!Tt{s}(!Tt{0})),q)!2!to !Tt{s}(!Tt{s}(!Tt{0})) !eq !Tt{s}(!Tt{0})!2$.
$!0!Tt{s}(!Tt{s}(!Tt{0})) !eq !Tt{s}(!Tt{0})$.
$!0!bot$.
!Econ{!0!1!lnot !It{square}(p) !eq !Tt{s}(!Tt{s}(!Tt{0}))*!It{square}(q)!2}
$!0!1!lnot !It{square}(p) !eq !Tt{s}(!Tt{s}(!Tt{0}))*!It{square}(q)!2$.
!Eass{!0!1!1!1!S !Tt{nat}(p)!2!land !1!S !Tt{nat}(q)!2!land !It{coprime}(p,q)!2!to !1!lnot !It{square}(p) !eq !Tt{s}(!Tt{s}(!Tt{0}))*!It{square}(q)!2!2}}
!Epr

!end
	
\end{plain}
\vspace{0.25cm}

Then we refine the previous attempt.
We define the properties we need for proving the main theorem
by \emph{stating} these auxiliary properties.
We do not prove \emph{any} of these properties for the moment.
For each proof, we just write: $\bot$ (\ie $\textit{false}$) is admitted (by \textbf{GAP}).
As from $\bot$ everything is true, each property holds.
This pseudo proof appears for the first lemma and is omitted for the others.

\begin{plain}

!lemma{nat:natsquare}{!0!1!all[n]!,!1(!1!S !Tt{nat}(n)!2!to !1!S !Tt{nat}(!It{square}(n))!2)!2!2}
!Pr{
$!0!bot$! by~{!bf GAP}.}
!Epr

!lemma{twotimes:even}{!0!1!all[n,p]!,!1(!1!1!S !Tt{nat}(p)!2!land !Tt{s}(!Tt{s}(!Tt{0}))*p !eq n!2!to !1!S !Tt{even}(n)!2)!2!2}

!lemma{evenpp:evenp}{!0!1!all[p]!,!1(!1!S !Tt{even}(!It{square}(p))!2!to !1!S !Tt{even}(p)!2)!2!2}

!lemma{even:twotimes}{!0!1!all[n]!,!1(!1!S !Tt{even}(n)!2!to !1!ex[p]!,!1(!1!S !Tt{nat}(p)!2!land n !eq !Tt{s}(!Tt{s}(!Tt{0}))*p)!2!2)!2!2}

!lemma{npq:nnppqq}{!0!1!all[n,p,q]!,!1(n !eq p*q!to !It{square}(n) !eq !It{square}(p)*!It{square}(q))!2!2}

!lemma{sqr2:4}{!0!It{square}(!Tt{s}(!Tt{s}(!Tt{0}))) !eq !Tt{s}(!Tt{s}(!Tt{s}(!Tt{s}(!Tt{0}))))}

!lemma{simplify:by2}{!0!1!all[n,p]!,!1(!Tt{s}(!Tt{s}(!Tt{s}(!Tt{s}(!Tt{0}))))*n !eq !Tt{s}(!Tt{s}(!Tt{0}))*p!to !Tt{s}(!Tt{s}(!Tt{0}))*n !eq p)!2!2}

!lemma{even:div2}{!0!1!all[n]!,!1(!1!S !Tt{even}(n)!2!to !It{divisor}(!Tt{s}(!Tt{s}(!Tt{0})),n))!2!2}

!end
	
\end{plain}
\vspace{0.35cm}

Assuming for the moment that these properties are true,
here is the new version of the main theorem.
It mimics the informal proof we started with,
but with a bit more details so that there is no gap.
Gaps now only appear in the lemmas. The file is proof-checked. 

\begin{plain}

!theorem{sqrt2:irrational}{!0!1!all[p,q]!,!1(!1!1!S !Tt{nat}(p)!2!land !1!S !Tt{nat}(q)!2!land !It{coprime}(p,q)!2!to !1!lnot !It{square}(p) !eq !Tt{s}(!Tt{s}(!Tt{0}))*!It{square}(q)!2)!2!2}
!Pr{
!Ass{!0!1!1!S !Tt{nat}(p)!2!land !1!S !Tt{nat}(q)!2!land !It{coprime}(p,q)!2}
!Con{!0!It{square}(p) !eq !Tt{s}(!Tt{s}(!Tt{0}))*!It{square}(q)}
$!0!1!S !Tt{nat}(!It{square}(q))!2$!by{Lemma}{nat:natsquare}.
$!0!Tt{s}(!Tt{s}(!Tt{0}))*!It{square}(q) !eq !It{square}(p)$.
$!0!1!S !Tt{even}(!It{square}(p))!2$!by{Lemma}{twotimes:even}.
$!0!1!S !Tt{even}(p)!2$!by{Lemma}{evenpp:evenp}.
$!0!1!ex[r]!,!1(!1!S !Tt{nat}(r)!2!land p !eq !Tt{s}(!Tt{s}(!Tt{0}))*r)!2!2$!by{Lemma}{even:twotimes}.
!Ex[r]{!0!1!1!S !Tt{nat}(r)!2!land p !eq !Tt{s}(!Tt{s}(!Tt{0}))*r!2}
$!0!It{square}(p) !eq !It{square}(!Tt{s}(!Tt{s}(!Tt{0})))*!It{square}(r)$!by{Lemma}{npq:nnppqq}.
$!0!It{square}(!Tt{s}(!Tt{s}(!Tt{0}))) !eq !Tt{s}(!Tt{s}(!Tt{s}(!Tt{s}(!Tt{0}))))$!by{Lemma}{sqr2:4}.
$!0!It{square}(p) !eq !Tt{s}(!Tt{s}(!Tt{s}(!Tt{s}(!Tt{0}))))*!It{square}(r)$.
$!0!Tt{s}(!Tt{s}(!Tt{s}(!Tt{s}(!Tt{0}))))*!It{square}(r) !eq !Tt{s}(!Tt{s}(!Tt{0}))*!It{square}(q)$.
$!0!Tt{s}(!Tt{s}(!Tt{0}))*!It{square}(r) !eq !It{square}(q)$!by{Lemma}{simplify:by2}.
$!0!1!S !Tt{nat}(!It{square}(r))!2$!by{Lemma}{nat:natsquare}.
$!0!1!S !Tt{even}(!It{square}(q))!2$!by{Lemma}{twotimes:even}.
$!0!1!S !Tt{even}(q)!2$!by{Lemma}{evenpp:evenp}.
!Eex{!0!1!S !Tt{even}(q)!2}
$!0!It{divisor}(!Tt{s}(!Tt{s}(!Tt{0})),q)$!by{Lemma}{even:div2}.
$!0!It{divisor}(!Tt{s}(!Tt{s}(!Tt{0})),p)$!by{Lemma}{even:div2}.
$!0!1!S !Tt{nat}(!Tt{s}(!Tt{s}(!Tt{0})))!2$.
$!0!It{coprime}(p,q)$.
$!0!1!all[r]!,!1(!1!1!S !Tt{nat}(r)!2!land !It{divisor}(r,p)!land !It{divisor}(r,q)!2!to r !eq !Tt{s}(!Tt{0}))!2!2$!by{Definition}{coprime/2}.
$!0!1!1!1!S !Tt{nat}(!Tt{s}(!Tt{s}(!Tt{0})))!2!land !It{divisor}(!Tt{s}(!Tt{s}(!Tt{0})),p)!land !It{divisor}(!Tt{s}(!Tt{s}(!Tt{0})),q)!2!to !Tt{s}(!Tt{s}(!Tt{0})) !eq !Tt{s}(!Tt{0})!2$.
$!0!Tt{s}(!Tt{s}(!Tt{0})) !eq !Tt{s}(!Tt{0})$.
$!0!bot$.
!Econ{!0!1!lnot !It{square}(p) !eq !Tt{s}(!Tt{s}(!Tt{0}))*!It{square}(q)!2}
$!0!1!lnot !It{square}(p) !eq !Tt{s}(!Tt{s}(!Tt{0}))*!It{square}(q)!2$.
!Eass{!0!1!1!1!S !Tt{nat}(p)!2!land !1!S !Tt{nat}(q)!2!land !It{coprime}(p,q)!2!to !1!lnot !It{square}(p) !eq !Tt{s}(!Tt{s}(!Tt{0}))*!It{square}(q)!2!2}}
!Epr

!end
	
\end{plain}
\vspace{0.35cm}

For each lemma, an LPTP warning is reported:
\texttt{there is a gap in the proof.}
But our initial draft 
is syntactically and logically coherent.
Moreover, the level of granularity  seems reasonable, 
at least for us. We moved from the informal usual high-level proof 
of the beginning of this section
to  the LPTP equivalent of the low-level Peano axioms.
Note that even the most elementary pieces (like $2^2 = 4$) 
of the above reasoning have to be proven (see Lemma \emph{sqr2:4}).
Let us ask for help for proving these lemmas by invoking an LLM.

\section{The proof completed}
\label{The-proof-completed}
We choose Claude from Anthropic and do the experiment with
Opus 4.5, the most powerful large language model released by Anthropic in late 2025. 
As this model is well known for its efficiency in programming tasks,
we expect it to be able to construct proofs validated with the LPTP proof checker.
We did not try any cheaper model (but we also run resolution-based automated theorem provers).
All interactions used Anthropic's default decoding settings (no temperature or sampling tuning).
We perform \textit{in-context learning}, \ie we feed the LLM with the LPTP user manual in PDF format.
We also give Claude two basic LPTP libraries (Prolog code and LPTP proofs in text format), one for Peano numbers and the other one about lists. The code and library of properties
for Peano numbers will be used in the proofs 
generated by Claude. On the other hand, the code and 
properties for lists will not be directly used but
give many other proof examples to Claude.

Our interaction follows a simple feedback loop: we submit a lemma
 (most often in LPTP syntax, sometimes in natural language) to Claude, feed its proof to
 the LPTP proof checker, and on failure forward to Claude the first
 \texttt{incorrect derivation step} reported by LPTP. We iterate until
 either LPTP accepts the proof (\DP{} or \BF{} in the tables below)
 or we estimate that Claude is not converging, in which case we provide
 a natural-language hint (\PH{}) or the proof itself (\PG{}).
 The human developer is responsible for choosing the
  order in which lemmas are attempted and for deciding when to abandon
  a back-and-forth. The ability of LPTP to localize the first incorrect
  step is what makes this loop tractable in practice.

So we start by asking the proofs of the easiest 
(from our point of view) lemmas, one lemma after the other.
We state the lemmas formally using the LPTP syntax.
The proofs  of lemmas \textit{nat:natsquare} 
(informally: $\mathit{nat}(n) \rightarrow \mathit{nat}(n^2)$), \textit{sqr2:4} ($2^2 = 4$),
\textit{twotimes:even} ($n = 2p \rightarrow \mathit{even}(n)$), \textit{even:twotimes}
($\mathit{even}(n) \rightarrow n=2p)$), and \textit{even:div2} ($\mathit{even}(n) \rightarrow 2 | n$)
are proposed by Claude and accepted by the proof checker of LPTP. 
For instance, our prompt for lemma \textit{sqr2:4} reads:                                        
  \begin{quote}
    \texttt{Donne-moi une preuve LPTP de :}\\
    \texttt{:- lemma(sqr2:4, square(s(s(0))) = s(s(s(s(0)))), ff by gap).}
  \end{quote}
  mixing a brief French instruction with the formal LPTP statement, the
  \texttt{ff by gap} clause being the placeholder to be filled in.
With each of its proofs, Claude summarizes in natural language
the idea of the proof.
For lemma \textit{sqr2:4}, Claude 
explains that it first shows that 0, 1 and 2 are natural numbers.
Then it computes $\mathit{times}(2,2,4)$ step by step: 
we have $\mathit{times}(0,2,0)$, $\mathit{plus}(2,0,2)$ hence $\mathit{times}(1,2,2)$.
We also have $\mathit{plus}(2,2,4)$ hence $\mathit{times}(2,2,4)$.
Finally, we get $\mathit{square}(2) = 4$.

The proofs of \textit{twotimes:even} ($2p = n \rightarrow \mathit{even}(n)$)
and its reciprocal \textit{even:twotimes} rely heavily on results from the
$\mathit{nat}$ library given to Claude initially, which it uses adequately.

For the lemma \textit{simplify:by2} ($4n = 2p \rightarrow 2n = p$),
we have to handle a few back and forths between Claude
and LPTP. 
The proof checker detects some 
\textit{incorrect derivation steps} at some places 
in the generated proof.
Each time, Claude analyzes the error and
proposes a new proof. 
Claude proposed and proved the auxiliary
lemma \textit{times:injective:second}.
Finally, Claude was able to
correct all the errors.

We observe a similar behaviour for 
the lemma \textit{npq:nnppqq} ($ n=pq \rightarrow n^2 = p^2 q^2$).

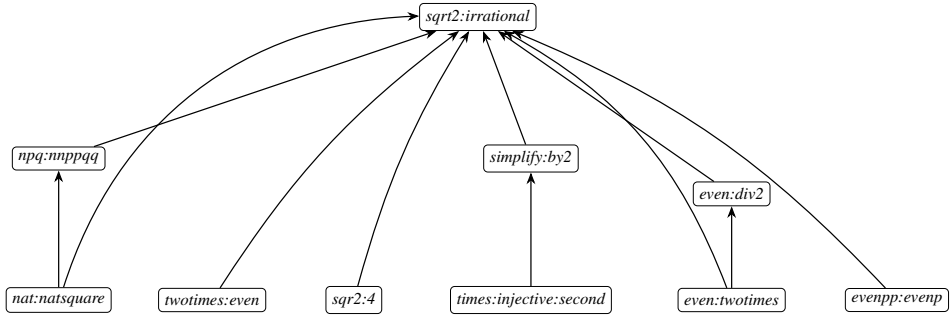
\begin{figure}[htbp]
\centering
\scalebox{0.78}{%
\begin{tikzpicture}[
  >=Stealth,
  box/.style={draw, rounded corners=2pt, font=\scriptsize\itshape, inner sep=3pt},
  arr/.style={->, semithick}
]

%% ===== Layer 0: leaves of Table~1 =====
\node[box] (nns)  at (0,    0) {nat:natsquare};
\node[box] (tte)  at (2.6,  0) {twotimes:even};
\node[box] (s24)  at (5.0,  0) {sqr2:4};
\node[box] (tis)  at (8.0,  0) {times:injective:second};
\node[box] (et2)  at (11.4, 0) {even:twotimes};
\node[box] (epe)  at (14.2, 0) {evenpp:evenp};

%% ===== Layer 1: intermediate lemmas =====
\node[box] (npq)  at (0,    2.4) {npq:nnppqq};
\node[box] (sby)  at (8.0,  2.4) {simplify:by2};
\node[box] (ed2)  at (11.4, 1.8) {even:div2};

%% ===== Layer 2: main theorem =====
\node[box] (sqrt2) at (7.1, 4.8) {sqrt2:irrational};

%% ===== Edges within Table~1 =====
\draw[arr] (nns) -- (npq);
\draw[arr] (tis) -- (sby);
\draw[arr] (et2) -- (ed2);

%% ===== Edges to sqrt2:irrational =====
\draw[arr] (npq) -- (sqrt2);
\draw[arr] (sby) -- (sqrt2);
\draw[arr] (ed2) -- (sqrt2);
\draw[arr] (tte) to[bend left=10]  (sqrt2);
\draw[arr] (s24) to[bend left=8]   (sqrt2);
\draw[arr] (epe) to[bend right=12] (sqrt2);
\draw[arr] (nns) to[bend left=35]  (sqrt2);
\draw[arr] (et2) to[bend right=20] (sqrt2);

\end{tikzpicture}}
\caption{Dependency graph of the lemmas of Table~\ref{tab:results1}.
  An arrow $A\!\to\!B$ means $A$ is used in the proof of $B$.}
\label{fig:dag1}
\end{figure}

Table~\ref{tab:results1} gives our obtained results
and Figure~\ref{fig:dag1} shows the dependency graph of the proof.
We also add a column ATP for recording the
answers we  get from running two automated theorem provers
on the same lemmas within a 20 seconds timeout.
This approach is fully described in \cite{MesnardMP24}
and summarized in the next section.

\begin{table}[hbtp]
    \centering  
\renewcommand{\arraystretch}{1.2}
\begin{tabular}{|l|l|c|c|}
\hline
Lemma or theorem & Shorthand & LLM & ATP \\
\hline
\hline
\textit{nat:natsquare} 
& $\mathit{nat}(n)\rightarrow \mathit{nat}(\mathit{square}(n))$
& \DP
& yes \\
\hline
\textit{twotimes:even} 
& $\mathit{nat}(p)\wedge 2\times p=n\rightarrow \mathit{even}(n)$
& \DP
& no \\
\hline
\textit{evenpp:evenp} 
& $\mathit{even}(\mathit{square}(p))\rightarrow \mathit{even}(p)$
& \NP
& no \\
\hline
\textit{even:twotimes} 
& $\mathit{even}(n) \rightarrow \exists p: \mathit{nat}(p)\wedge n = 2\times p$
& \DP 
& no \\
\hline
\multirow{2}{*}{\textit{npq:nnppqq}} 
& $n=p\times q\rightarrow$

& \multirow{2}{*}{\BF} 
& \multirow{2}{*}{no}\\
& \quad $\mathit{square}(n)=\mathit{square}(p)\times\mathit{square}(q)$
&
& \\
\hline
\textit{sqr2:4} 
& $\mathit{square}(2)=4$
& \DP 
& yes \\
\hline
\textit{times:injective:second} 
& $ (n+1)\times p=(n+1)\times q \rightarrow p = q $
& \DP 
& yes \\
\hline
\textit{simplify:by2} 
& $4\times n=2\times p\rightarrow 2\times n =p$
& \BF 
& no \\
\hline
\textit{even:div2} 
& $\mathit{even}(n)\rightarrow \mathit{divisor}(2,n)$
& \DP 
& yes \\
\hline
\multirow{2}{*}{\textit{sqrt2:irrational}} 
& $\mathit{nat}(p)\wedge \mathit{nat}(q)\wedge \mathit{coprime}(p,q)\rightarrow$
& \multirow{2}{*}{\PG} 
& \multirow{2}{*}{yes} \\
& \quad $\neg \mathit{square}(p)=2\times\mathit{square}(q)$ & & \\
\hline
\end{tabular}
\caption{Summary of the obtained proof results. Shorthands for the column LLM: \DP:  directly proven by Claude, \NP: not proven by Claude, \BF: proven by Claude with back and forths, and \PG: proof was given to Claude. A \textit{yes} 
in the ATP column means a proof was found by one of our two provers.}
\label{tab:results1}
\end{table}

The last remaining property, \textit{evenpp:evenp}
($\mathit{even}(p^2) \rightarrow \mathit{even}(p)$), proved much trickier:
Claude failed to find a proof or even go in the right direction.
So we derived the main steps by hand and then asked Claude %bottom-up
to prove all auxiliary lemmas stated in natural language.

%The last remaining property to be proven happens to be much trickier. 
%The lemma \textit{evenpp:evenp} states that 
%$\mathit{even}(p^2) \rightarrow \mathit{even}(p)$.
%Claude was not able to find a proof nor to 
%go in the right direction. 
%So  we first derive the main steps of the proof by hand.
%Then going bottom-up this time,
%we decide to state all the auxiliary
%lemmas in natural language, asking Claude to prove them.
\begin{table}[hbtp]
\begin{center}
\renewcommand{\arraystretch}{1.2}
\begin{tabular}{|l|l|c|c|}
    \hline
    Lemma or theorem & Shorthand & LLM & ATP \\
    \hline
    \hline
    \textit{even:succ:odd} &
        $\mathit{even}(p) \rightarrow \mathit{odd}(p+1)$ 
        & \DP 
        & yes \\
    \hline
    \textit{odd:succ:even} &
        $\mathit{odd}(p) \rightarrow \mathit{even}(p+1)$ 
        & \DP
        & yes \\
    \hline
    \textit{nat:even:disj:odd} &
        $\mathit{nat}(p) \rightarrow \mathit{even}(p) \lor \mathit{odd}(p)$ 
        & \DP
        & yes \\
    \hline
%    \textit{nat:even:termination} &
%        $\mathit{nat}(n) \rightarrow \mathit{even}(n) \ \mathit{terminates} \land \mathit{even}(n+1) 
%        \ \mathit{terminates}$  
%        & \DP
%        & yes \\
%    \hline
%    \textit{even:termination} &
%        $\mathit{nat}(n) \rightarrow \mathit{even}(n) \ \mathit{terminates}$ 
%        & \DP
%        & yes \\
%    \hline
%    \textit{nat:odd:termination}
%    & $\mathit{nat}(n) \rightarrow \mathit{odd}(n) \ \mathit{terminates} \land \mathit{odd}(n+1) 
%        \ \mathit{terminates}$   
%    & \DP
%    & yes \\
%    \hline
%    \textit{odd:termination}
%    & $\mathit{nat}(n) \rightarrow \mathit{odd}(n) \ \mathit{terminates}$ 
%    & \DP
%    & yes \\
%    \hline
%    \textit{even:negation:odd} &
%        $\mathit{even}(n) \rightarrow \mathit{odd}(n)\ \mathit{fails}$ 
%        & \DP
%        & yes \\
%    \hline
    \textit{odd:negation:even} &
        $\mathit{odd}(n) \rightarrow \mathit{even}(n)\ \mathit{fails}$ 
        & \DP
        & yes \\
    \hline
    \textit{even:types}
    & $\mathit{even}(n)\rightarrow \mathit{nat}(n)$
    & \DP
    & yes \\
    \hline
    \textit{odd:types}
    & $\mathit{odd}(n)\rightarrow \mathit{nat}(n)$
    & \DP
    & yes \\
    \hline
    \textit{odd:plus:odd}
    & $\mathit{odd}(m)\wedge \mathit{odd}(n) \rightarrow \mathit{even}(m+n)$
    & \DP
    & no \\
    \hline
    \textit{even:plus:odd}
    & $\mathit{even}(m)\wedge \mathit{odd}(n) \rightarrow \mathit{odd}(m+n)$
    & \DP
    & yes \\
    \hline
    \textit{odd:times:odd}
    & $\mathit{odd}(m)\wedge \mathit{odd}(n) \rightarrow \mathit{odd}(m\times n)$
    & \DP
    & no \\
    \hline
    \textit{odd:square:odd} &
        $\mathit{nat}(p) \land \mathit{odd}(p) \rightarrow \mathit{odd}(p^2)$ 
        & \DP
        & yes \\
    \hline
    \textit{evenpp:evenp}
    & $\mathit{nat}(p)\wedge\mathit{even}(\mathit{square}(p)) \rightarrow \mathit{even}(p)$
    & \PH
    & yes \\
    \hline
    \multirow{2}{*}{\textit{sqrt2:irrational}} 
    & $\mathit{nat}(p)\wedge \mathit{nat}(q)\wedge \mathit{coprime}(p,q)\rightarrow$
    & \multirow{2}{*}{\PG} 
    & \multirow{2}{*}{no} \\
    & \quad $\neg \mathit{square}(p)=2\times\mathit{square}(q)$ & & \\
\hline
    
\end{tabular}
\end{center}
\caption{Summary of the obtained proof results for auxiliary lemmas: \DP:  directly proven by Claude, \PH: proven by Claude with given hints, and \PG: proof was given to Claude}
\label{tab:results2}
\end{table}

 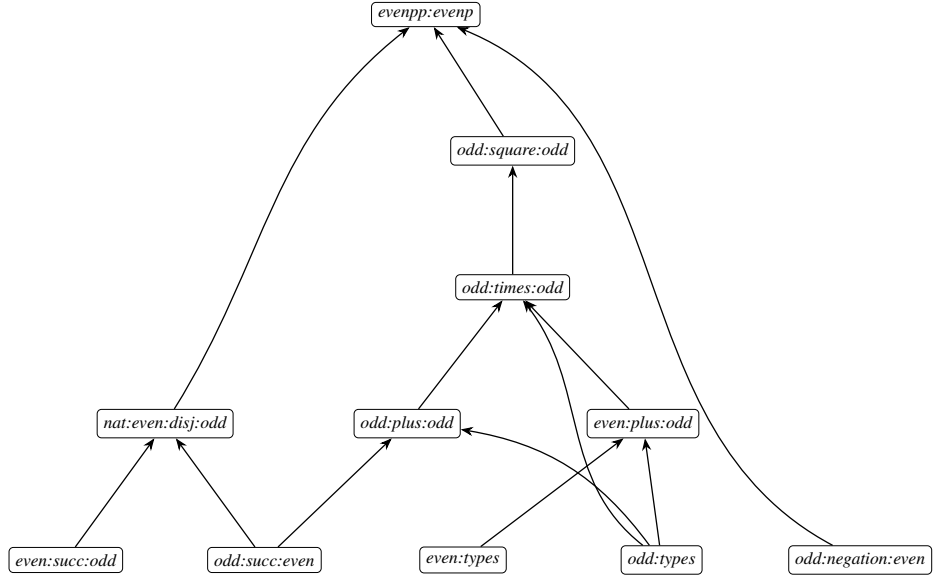
\begin{figure}[htbp]
\centering
\scalebox{0.82}{%
\begin{tikzpicture}[
  >=Stealth,
  box/.style={draw, rounded corners=2pt, font=\scriptsize\itshape, inner sep=3pt},
  arr/.style={->, semithick}
]

%% ===== Layer 0: feuilles =====
\node[box] (esucc)  at (0,    0) {even:succ:odd};
\node[box] (osucc)  at (3.2,  0) {odd:succ:even};
\node[box] (etypes) at (6.4,  0) {even:types};
\node[box] (otypes) at (9.6,  0) {odd:types};
\node[box] (onege)  at (12.8, 0) {odd:negation:even};

%% ===== Layer 1 =====
\node[box] (nedo) at (1.6, 2.2) {nat:even:disj:odd};
\node[box] (opo)  at (5.5, 2.2) {odd:plus:odd};
\node[box] (epo)  at (9.3, 2.2) {even:plus:odd};

%% ===== Layer 2 =====
\node[box] (oto)  at (7.2, 4.4) {odd:times:odd};

%% ===== Layer 3 =====
\node[box] (oso)  at (7.2, 6.6) {odd:square:odd};

%% ===== Layer 4 =====
\node[box] (epe)  at (5.8, 8.8) {evenpp:evenp};

%% ===== Layer 5 =====
%\node[box] (sqrt2) at (5.8, 11.0) {sqrt2:irrational};

%% ===== Arêtes =====
\draw[arr] (esucc)  -- (nedo);
\draw[arr] (osucc)  -- (nedo);
\draw[arr] (osucc)  -- (opo);
\draw[arr] (otypes) to[bend right=22] (opo);
\draw[arr] (etypes) -- (epo);
\draw[arr] (otypes) -- (epo);
\draw[arr] (otypes) to[out=142, in=308] (oto);
\draw[arr] (opo)    -- (oto);
\draw[arr] (epo)    -- (oto);
\draw[arr] (oto)    -- (oso);
\draw[arr] (nedo)   to[out=58, in=218] (epe);
\draw[arr] (oso)    -- (epe);
\draw[arr] (onege)  to[out=152, in=338] (epe);
%\draw[arr] (epe)    -- (sqrt2);

\end{tikzpicture}}
\caption{Dependency graph of the auxiliary lemmas of Table~\ref{tab:results2}.
  An arrow $A\!\to\!B$ means $A$ is used in the proof of $B$.
%  \textit{sqrt2:irrational} additionally uses lemmas from Table~1 (not shown).
}
\label{fig:dag}
\end{figure}

Finally we ask Claude to prove the lemma
\textit{evenpp:evenp} ($\mathit{nat}(p) \land \mathit{even}(p^2) \rightarrow \mathit{even}(p)$)
using the following hint. 
\begin{quote}
Either $p$ is even or odd. If $p$ is even, we are done.
Otherwise $p$ is odd so $p^2$ is odd, hence $\mathit{even}(p^2)$
is false. Thus $\mathit{nat}(p) \land \mathit{even}(p^2)$ is false and the implication is true. So in all cases, the implication is true.
\end{quote}
Claude generates the proof following our suggestion.
Finally LPTP validates the whole proof file.
Note that the ATPs were not able to prove the main theorem, 
contrary to the previous experiment. It is likely that 
they need more time to deal with a larger search space due to 
the increased number of lemmas. Table \ref{tab:results2}  presents
the results and Figure~\ref{fig:dag} shows the dependency graph of the proof.

\section{Back to Prolog}
\label{Back-to-Prolog}

We now have an LPTP proof that $\sqrt{2}$ is irrational.
So on a conceptual level, we know that solving the Prolog query
of Section \ref{Introduction} is hopeless. Within the LPTP framework, 
let's take a closer look  by first introducing the Prolog rule:
\begin{verbatim}
sqrt2_is_rational :-    
    nat(S),plus(P,Q,S),gcd(P,Q,s(0)),times(P,P,P2),times(Q,Q,Q2),plus(Q2,Q2,P2).
\end{verbatim}
In the meantime, Opus 4.6 is out (released early February 2026), and
we switch to this new version in \emph{Code mode}.
We give Claude access to the \emph{full Prolog code} of LPTP.
Contrary to the \emph{Chat mode}, Claude is able to call
the proof checker itself, and we allow this capability.
We want to prove an operational property of our Prolog code,
so we create a \texttt{CLAUDE.md} file containing
the following \textit{Problem specification}:
\begin{quote} 
The predicate `sqrt2_is_rational/0' is defined in `sqrt2.pl'. This file also uses predicates from the files `nat.pl' and `gcd.pl' present in the directory. The file `sqrt2_v2_no_gap.pr' contains a valid  LPTP proof that the square root of 2 is not a rational. This proof should help to prove with LPTP  that the query `?- sqrt2_is_rational.' does not succeed.
\end{quote}

Furthermore, we point out %to Claude 
that a bridge is needed between
the logical definition of \texttt{coprime(p,q)} used
in the hypothesis of Theorem [\textit{sqrt2:irrational}]
and the condition \texttt{gcd(P,Q,s(0))} appearing in the clause above
(actually, the conditions are equivalent). 
Claude presents a first plan.
Initially it wants to prove that 
$\mathit{fails} \ sqrt2\_is\_rational$.
We explain that, as termination of \texttt{sqrt2\_is\_rational}
can not be established, we do not have the usual dichotomy between 
$\mathit{success}$ and $\mathit{failure}$ (Axiom 7 of Section 2). 
The best we can hope is a 
proof of $\neg \ \mathit{succeeds} \ sqrt2\_is\_rational$ (Axiom 6).
As Claude  has  access to the Prolog code of LPTP,
it does use this access to have a better understanding
of the LPTP syntax and of the LPTP proof checking process.
Claude is able to generate three bridging lemmas and to prove
that \texttt{sqrt2\_is\_rational} can not succeed by contradiction.

% Table \ref{tab:results3} presents the results.
%
% \begin{table}[hbtp]
% \begin{center} 
% \renewcommand{\arraystretch}{1.2}
% \begin{tabular}{|l|l|c|c|}
% \hline
% Lemma or theorem & Shorthand & LLM & ATP \\
% \hline
% \hline
% \textit{divisor:of:one} 
% & $\mathit{nat}(r) \wedge \mathit{divisor}(r,s(0)) \rightarrow r=s(0)$
% & \DP
% & yes \\
% \hline
% \textit{gcd:one:coprime} 
% & $\mathit{nat}(p)\wedge \mathit{nat}(q) \wedge \mathit{gcd}(p,q,s(0))
% \rightarrow \mathit{coprime}(p,q)$
% & \DP
% & no \\
% \hline
% \textit{plus:double:twotimes} 
% & $\mathit{nat}(p) \wedge \mathit{plus}(p,p,q) \rightarrow q=s(s(0))\times p$
% & \DP
% & yes \\
% \hline
% \textit{sqrt\_two:irrational:query} 
% & $\neg \mathit{succeeds} (\mathit{sqrt2\_is\_irrational})$
% & \DP 
% & yes \\
% \hline
% \end{tabular}
% \end{center}
% \caption{Summary of the obtained proof results. Shorthands for the column LLM: \DP:  directly proven by Claude. A \textit{yes} in the ATP column means a proof was found by one of our two provers.}
% \label{tab:results3}
% \end{table}

\section{Related Work}
\label{Related-work}

There are a few Prolog verification frameworks, see, \eg  \cite{Apt94b,Deransart93a,FerrandD93,PedreschiR99}
and more recently  \cite{Drabent16a}.
Most of them aim at \emph{paper-and-pencil} proofs
and do not provide any implemented proof checker.
Of course, as the most recent paper was published ten years ago,
none of these papers reports 
any form of interaction with an LLM.

% A summary of \cite{2006provers}: TBD - EP
A comparison of seventeen proof assistants used in formal mathematics is provided in~\cite{2006provers}
through their formalization of a proof of the irrationality of $\sqrt{2}$. 
A proof in each system is presented, highlighting differences in logical foundations, syntax, automation, and usability. The main goal of the book is to illustrate the diversity of formal reasoning systems and to evaluate the state of proof assistant technology in the early 2000s. The included systems satisfy two criteria: they are designed, or used, for the formalization of mathematics and they are better, in at least one dimension, than all other systems in the collection. We note that LPTP is not considered.
%and recall that its purpose is to prove
%Prolog programs properties in first order logic (FOL).
This work, published 20 years ago, does not of course report 
any interaction
with an LLM.

%ATP for LPTP: TBD - FM
Recently, we showed in \cite{MesnardMP26}
how one can easily plug any first order logic (FOL)
automated theorem prover (ATP) into LPTP.
We observe that inside the LPTP interactive development environment, namely Emacs,
the FOL axioms are hardwired within tactics.
Going back to the FOL formalization 
of the operational semantics of Prolog
described in \cite{Staerk98a}, and
translating the axioms in FOF, a human-readable syntax
for FOL~\cite{Sut23},
we can apply  any 
\emph{off-the-shelf} \ FOF-compatible ATP.
% \textbf{FOF?}\\
We applied this strategy to the whole LPTP library,
and obtained a success rate of about 80\%
with a 20 seconds timeout, running
two of the most successful ATPs.
The main advantage of this approach 
is that we can run the freely available ATPs 
locally
on our machines, without relying
on a non-free external provider.
Moreover, the experiments are easily reproducible.
On the other hand, we 
get  resolution proofs (more precisely superposition
calculus proofs) which
cannot  be directly rewritten as natural deduction
proofs as needed for LPTP. So we cannot 
proof-check these proofs with LPTP.
We experimented the approach on our $\sqrt{2}$ example.
The results appear in the ATP column of Tables~\ref{tab:results1}
and~\ref{tab:results2} and are summarized in the conclusion.

%LLM and formal reasoning: TBD - WV
Back to our case study, we have illustrated how a formal proof checker, LPTP, can be combined with an LLM (in our experiment Claude) in order to arrive at a successful (partial) automatization of a theorem proving process. In particular, we use the LLM to generate proof steps whose correctness is subsequently checked by the proof checker. This is known as proof-step generation (\eg~\cite{10.5555/3600270.3600878,polu2020generativelanguagemodelingautomated}) and contrasts somewhat with other approaches that try to generate the proof as a whole (\eg~\cite{10.1145/3611643.3616243,jiang2023draftsketchproveguiding}).

Our observations are in line with the recent literature where similar combinations of language models and proof checkers are explored using a variety of approaches. Giving a complete overview of existing and ongoing work is outside the scope of the current paper, but we can cite the following.
For example, in \cite{xin2025deepseekproverv} a language model (DeepSeek-Prover-V1.5) is introduced that is specifically designed and trained for doing theorem proving in Lean 4. Being a model specifically trained on formal languages, it is in sharp contrast with other approaches using an off-the-shelf language model, including our own.  
In~\cite{thakur2024an}, the authors introduce COPRA a system that uses an off-the-shelf LLM (GPT-4) and in-context learning to let the model generate proof steps directly written in the formal language of Rocq or Lean. COPRA uses a backtracking search methodology to construct the proof in a step-by-step manner, using feedback from the proof checker to construct the prompt for the next step. 
Likewise,~\cite{teodorescu2024nlir} also uses a general-purpose model (GPT-4) but proposes to use natural language as an intermediate language. An interactive environment (called \textit{Pétanque}) provides two proof tactics that allow to convert the internal reasoning to Rocq code. 
In a somewhat similar way, the Hermes framework~\cite{ospanov-hermes-2025} couples informal reasoning by an LLM with a translator module that allows to translate this reasoning into Lean code and a prover module (such as Goëdel-Prover-V2) for verification.

\section{Conclusion}
\label{Conclusion}
We report in this paper  one of the very first interactions  
-- to the best of our knowledge --
between LPTP (Logic Program Theorem Prover) and an LLM.
We have chosen Claude from Anthropic and did the experiment with 
Opus 4.5. We fed the LLM with the LPTP user manual in PDF format 
and two basic LPTP libraries (Prolog code and LPTP proofs in text format), 
one for Peano numbers and the other one about lists.
We have selected a small classical  problem
among the proof assistant community,
namely the irrationality of $\sqrt 2$. This problem has been
the running example of a whole book \cite{2006provers}.
We did not find any LPTP proof of
the irrationality of $\sqrt 2$ on the Internet.

While we had to explicitly give the proof structure 
of the two main results -- namely \textit{sqrt2:irrational} and 
\textit{ evenpp:evenp} --,
%we were surprised to notice that 
the LLM was able 
to prove some simple properties, sometimes proposing
and proving new auxiliary lemmas. We stated the properties 
either in the LPTP language or in natural language (French in our case).
Sometimes after 
a few back and forths between the LLM and LPTP, but 
sometimes directly, the lemmas generated by Claude 
were approved by the proof checker.

Beyond the LLM, we also leveraged off-the-shelf first-order ATPs such as Vampire and E
by translating the LPTP axiomatization into FOF~\cite{Sut23,MesnardMP26}.
They proved 5 out of 10 lemmas in the intermediate version (Section~3)
and 9 out of 12 in the final, finer-grained version (Section~4).
ATPs and the LLM play complementary roles: ATPs are free, local, fast and reproducible,
but their resolution-based proofs cannot be easily rewritten as natural deduction
and thus cannot directly enrich the LPTP library. 
%--- for that,
%an LLM-assisted, LPTP-checked derivation remains necessary.

As already implied by the literature (\eg~\cite{LPAR2024:Automated-Theorem-Provers-Help,thakur2024an}), 
the proof checker is the key element 
to eradicate LLM's hallucinations.
It allows the automatic and efficient 
classification of the LLM's answers:
either \emph{correct} -- the proof checker validates the proposed proof --
or \emph{wrong} -- the proof checker does not validate the proposed proof and reports the first \emph{incorrect derivation step} within the proposed proof --. The ability to point out the first step 
in the proof that is not correct is a great help for the LLM.
Note also that once a proof is checked, the corresponding result
can be safely added to the LPTP library, without compromising its
logical consistency. 
This last point is crucial as
the introduction of a false result in the library 
would allow the proof of \textit{any} statement.

The interactions between LPTP, an LLM, and an LP developer
leading to a semi-LLM-generated LPTP
proof of the irrationality of $\sqrt 2$ is
the main contribution of this work\footnote{
The data from the experiment 
are available at \url{https://github.com/FredMesnard/LPTP-LLM.git}.}.
It paves the way to an AI-enhanced approach
for proving LP/Prolog properties,
as the relevance of LLMs will  continue to grow
in the future.
We plan to evaluate the combination LPTP/LLM
on a set  of Prolog verification problems.

%\vspace{0.1cm}
%\noindent
%\textbf{Acknowledgement.} We thank Romain Pabot 
%for enlightening discussions about LLMs.

%\newpage
\bibliographystyle{eptcs}
\bibliography{biblio.bib,wim}

%\newpage 
%\section*{For the reviewers: the final proof}
%\begin{plain}
%\input{sqrt2_final_no_gap.tex}	
%\end{plain}

\end{document}